\documentclass[aps, reprint, amsmath,amssymb, pra, superscriptaddress]{revtex4-2}
\usepackage{svg}
\usepackage{color}
\usepackage{graphicx}
\usepackage{times}
\usepackage{dcolumn}
\usepackage{bm}
\usepackage{physics}
\usepackage{hyperref}
\usepackage{url}
\usepackage{amsmath}
\usepackage{tikz-cd}
\usepackage{notes2bib}
\usepackage{multirow}
\usepackage{mathrsfs}
\usepackage{tabularx}
\usepackage{ dsfont }
\usepackage{svg}
\usepackage{verbatim}
\usepackage{amssymb}
\usepackage{amsfonts}              
\usepackage{amsthm}                
\usepackage{mathtools}
\usepackage{qcircuit}
\usepackage[normalem]{ulem}
\usepackage{xcolor}
\usepackage{csquotes}
\usepackage[many]{tcolorbox}
\usetikzlibrary{arrows.meta, calc}

\usepackage[clock]{ifsym}
\usepackage{fontawesome}
\usepackage{manfnt}

\usepackage{enumitem}

\DeclarePairedDelimiterX{\barpair}[2]{(}{)}{%
  #1\;\delimsize\|\;#2%
}

\usepackage{stackengine,scalerel}
\newcommand\dddag{%
  {\sbox0{\ddag}\scalerel*{%
  \stackengine{-.6\ht0}{\ddag}{\ddag}{O}{c}{F}{F}{S}}{\ddag}}%
}

\theoremstyle{definition}
\newtheorem{theorem}{Theorem}
\newtheorem{prop}{Proposition}
\newtheorem{remark}{Remark}

\newtheorem{lemma}{Lemma}

\newtheorem{definition}{Definition}
\newtheorem{corollary}{Corollary}

\newcommand{\Ad}{\text{Ad}}
\newcommand{\id}{\text{id}}
\newcommand{\mf}[1]{\mathfrak{#1}}
\newcommand{\mcal}[1]{\mathcal{#1}}
\newcommand{\mscr}[1]{\mathscr{#1}}

\newcommand{\jami}{Jamio{\l}kowski }

\newcommand{\mds}[1]{\mathds{#1}}

\usepackage[normalem]{ulem}
\bibnotesetup{note-name    = ,use-sort-key = false}
\begin{document}

\title{Uniqueness of quantum state over time function}

\author{Seok Hyung Lie}
\affiliation{
 School of Physical and Mathematical Sciences, Nanyang Technological University, 21 Nanyang Link, Singapore, 637371
}%
\author{Nelly H.Y. Ng}
\affiliation{
 School of Physical and Mathematical Sciences, Nanyang Technological University, 21 Nanyang Link, Singapore, 637371
}%

\date{\today}

\begin{abstract}
A fundamental asymmetry exists within the conventional framework of quantum theory between space and time, in terms of representing causal relations via quantum channels and acausal relations via multipartite quantum states. Such a distinction does not exist in classical probability theory. In effort to introduce this symmetry to quantum theory, a new framework has recently been proposed, such that dynamical description of a quantum system can be encapsulated by a static \emph{quantum state over time}. 
In particular, Fullwood and Parzygnat recently proposed the state over time function based on the Jordan product as a promising candidate for such a quantum state over time function, by showing that it satisfies all the axioms required in the no-go result by Horsman et al. However, it was unclear if the axioms induce a unique state over time function.
In this work, we demonstrate that the previously proposed axioms cannot yield a unique state over time function. In response, we therefore propose an alternative set of axioms that is operationally motivated, and better suited to describe quantum states over any spacetime regions beyond two points. By doing so, we establish the Fullwood-Parzygnat state over time function as the essentially unique function satisfying all these operational axioms.
\end{abstract}

\pacs{Valid PACS appear here}
\maketitle

\textit{Introduction.}--- Oftentimes, quantum theory has been considered a generalization of the classical probability theory in the sense it is essentially a theory for calculating probabilities of measurement outcomes \cite{debrota2021born}. However, unlike the classical theory, there  is a fundamental asymmetry between space and time in the conventional formalism of quantum theory. In the classical theory, both time-like and space-like correlations can be described with joint distributions, but in quantum theory, the space-like correlation can be expressed as a multi-partite quantum state, while the time evolution should be described with quantum channels. Is it really impossible to formalize a ``causally neutral" quantum theory \cite{leifer2013towards}? There have been various attempts to solve this problem \cite{feynman1948space,sorkin1997forks,griffiths1984consistent,omnes1988logical,aharonov2008two,aharonov2009multiple,chiribella2012perfect,chiribella2013quantum,oreshkov2012quantum,costa2016quantum,hardy2007towards,hardy2012operator,cotler2018superdensity,cotler2019quantum} and a particular effort lies in mapping dynamical quantum processes to static quantum state over time, so that every correlation can be expressed as a quantum state regardless of their causal structure. 

There have been multiple proposals for the candidates of state over time functions (or quantum Bayes maps that could be related to states over time through the result of Ref. \cite{parzygnat2023time}) including those by Ohya \cite{ohya1983note}, Leifer-Spekkens (LS) \cite{leifer2013towards}, Wilde \cite{wilde2015recoverability}, Fitzsimons-Jones-Vedral (FJV) \cite{fitzsimons2015quantum} and Sutter-Tomamichel-Harrow (STH) \cite{sutter2016strengthened}. (See Ref. \cite{parzygnat2023time} for a comprehensible introduction to state over time functions and their applications to the quantum Bayes' rule.) A no-go result by Horsman et al. \cite{horsman2017can} seemingly forbids the existence of state over time functions that satisfy several mathematical axioms, but Fullwood and Parzygnat circumvented this \cite{fullwood2022quantum} by appropriately adjusting the axioms to physically relevant forms, and introducing a new state over time function based on the Jordan product that is equivalent to the FJV function for qubits \cite{horsman2017can}.  However, whether the Fullwood-Parzygant function is the only one among its class was an open problem \cite{fullwood2022quantum,fullwood2023quantum}. In this work, we study the logical relation between different axioms that have been imposed on state over time function and show that the axioms employed previously are not strong enough to uniquely characterize a state over time function. We then propose a new set of axioms whose operational meaning is clearer than the previous one. Consequently, we show that the Fullwood-Parzygnat state over time function is the only one that satisfies all the axioms.

\textit{Notations.---}
\begin{table}[]
    {\small
    \begin{tabularx}{0.95 \linewidth}{ |c|X| } 
 \hline
 $A,B,\cdots$ & Systems and their associated Hilbert spaces. \\
 $\mf{S}(A)$  & Quantum states on system $A$. \\ 
 $\mf{C}(A,B)$  & Quantum channels from $A$ to $B$. If $A=B$, then we denote it by $\mf{C}(A)$.\\ 
 $\mf{B}(A,B)$ & Operators from Hilbert space $A$ to $B$. If $A=B$, we denote it by $\mf{B}(A)$.\\ 
 $\mf{H}(A)$  &  Hermitian operators on Hilbert space $A$.\\ 
 \hline
\end{tabularx}
}
    \caption{Symbols used in this paper.}
    \label{tab:notations}
\end{table}Table \ref{tab:notations} summarizes the common notations we adapt in this manuscript. We identify quantum systems and their associated Hilbert spaces and denote both by $A,B,\cdots$. The dimension of $A$ is denoted as $|A|:=\Tr[\mds{1}_A]$. The identity operator in $\mf{B}(A)$ is denoted by $\mds{1}_A$, and $\pi_A:=|A|^{-1}\mds{1}_A$ is the maximally mixed state on $A$. Similarly, if $A_0$ is a subspace of $A$, then $\mds{1}_{A_0}$ is understood both as the identity operator in $\mf{B}(A_0)$ and the orthogonal projector onto $A_0$ in $\mf{B}(A)$. To emphasize its domain and range, a quantum channel $\mcal{E}$ from $A$ to $B$ will be occasionally denoted by $\mcal{E}_{B|A}$. The identity channel between isomorphic systems $A$ and $B$ is denoted as $\id_{B| A}$ \footnote{Technically there cannot be ``the'' identity channel between different spaces $A$ and $B$, but we follow the convention in which isomorphic quantum systems are identified so that one can talk about same states in different spaces, e.g. $\ket{1}_A$ and $\ket{1}_B$ are the same states but in different spaces.}, and $\id_A:=\id_{A|A}$. For any $X,Y\in\mf{B}(A)$, $[X,Y]:=XY-YX$ is the commutator of $X$ and $Y$ and $\Ad_X(Y):=XYX^\dag$.

When we say that $\rho_{AB}$ is a quantum state over spacetime, it means either that $\rho_{AB}$ is a regular quantum state, i.e. $\rho_{AB} \geq 0$ and $\Tr\rho_{AB}=1$, or that $\rho_{AB}$ is a quantum state over time, depending on the state over time function of interest, i.e. $\rho_{AB}=\mcal{E}_{B|A}\star \rho_A$ for some $\mcal{E}_{B|A}$ and $\rho_A\in\mf{S}(A)$. (Definition of $\star$ is given in the subsequent section.) Without descriptions like `over (space-)time', a quantum state means the regular quantum state above. 

\textit{New axioms for state over time.}---  We start with a formal definition for the state over time function, as a mapping from a dynamical description of evolution of quantum systems to a static one, satisfying a set of minimal requirements.
\begin{definition} \label{def:star}
    A \textit{state over time function}, or a \textit{star product}, $\star: \mf{C}(A,B)\times \mf{S}(A) \to \mf{B}(AB)$ is a function that satisfies
    \begin{align}
        \Tr_A \mcal{E}_{B|A}\star \rho_A &= \mcal{E}_{B|A}(\rho_A), \label{eqn:marg}\\
        \Tr_B \mcal{E}_{B|A}\star \rho_A &= \rho_A. \label{eqn:marg2}
    \end{align}
Note that we can extend the domain to make $\star$ homogeneous in each argument, by letting $(\lambda\mcal{E}_{B|A})\star\rho_A=\mcal{E}_{B|A}\star(\lambda\rho_A)=\lambda
\left[\mcal{E}_{B|A}\star(\rho_A)\right]$ for all $\lambda\in\mds{C}$.
\end{definition}
 Furthermore, if $\star$ is linear in the first argument, we call it \textit{process-linear} and if it is linear in the second argument, we call it \textit{state-linear}. If it is both, then we call it \textit{bilinear}.
Although the trace of a quantum states over time should be normalized to one, it has been observed that such a state need \textit{not} be positive. 
This does not necessarily mean that its definition is pathological. As Fullwood described, it is akin to the negative sign associated with time in the spacetime interval in relativity \cite{fullwood2023quantum}. We carefully discuss the subtleties of linearity and non-positive properties in Appendix \ref{app:rmks}.
With this, we introduce the definition of the FP state over time function. 

\begin{definition} [\cite{fullwood2022quantum}]
    The Fullwood-Parzygnat state over time function $\star_{FP}$ (\textit{the FP function} in short) is defined for all linear maps $\mcal{E}$ from $\mf{B}(A)$ to $\mf{B}(B)$ as 
    \begin{equation}
        \mcal{E}_{B|A}\star_{FP}\rho_A:= \frac{1}{2} \left\lbrace \rho_A\otimes\mds{1}_B, \mscr{D}[\mcal{E}]\right\rbrace, 
    \end{equation}
    where $\lbrace X,Y \rbrace =XY+YX$ is the Jordan product (or anti-commutator) and $\mscr{D}[\mcal{E}]$ denotes the \textit{(Jamio{\l}kowski) channel state} of a channel $\mcal{E}\in\mf{C}(A,B)$ \cite{jamiolkowski1972linear}, defined as
\begin{equation}
    \mscr{D}[\mcal{E}]:=(\id_A\otimes \mcal{E}_{B|A'})\qty(F_{AA'}),
\end{equation}
and $F_{XY}$ denotes the swap gate between systems $X$ and $Y$ and $A'$ is a copy of $A$.
\end{definition}

The FP function has several useful mathematical properties, i.e. \textit{Hermiticity, bilinearity, preservation of classical limit} and \textit{associativity} (detailed explanation of the properties will be given in the following sections). These properties were proposed as axioms that should characterize any reasonable candidate for a state over time function \cite{horsman2017can,fullwood2022quantum,parzygnat2023time,fullwood2023quantum}.
However, it was an open problem if the FP function is the unique function satisfying such conditions \cite{fullwood2022quantum}. We solve this problem in Theorem \ref{thm:PF}. See Appendix \ref{App:B} for the proof. %

\begin{theorem}[Characterization of the FP function] \label{thm:PF}
    The Fullwood-Parzygnat (FP) state over time $\star_{FP}$ is the only state over time function that satisfies the axioms \textbf{(E), (P), (CC)} and \textbf{(T)} given below. 
\end{theorem}

In other words, the FP function is indeed unique when it comes to satisfying a closely related, but operationally motivated set of new axioms. These axioms are fomulated in consideration of multipartite settings, since a satisfactory state over time should be consistent with mixed causal structures~\cite{leifer2013towards}. Furthermore, by examining the relationship between our new axioms with the previous set, we will see later that the previous axioms are insufficient for inducing a unique state over time function. 

\begin{tcolorbox}[breakable, enhanced jigsaw, colback=white, colframe=black]
\textbf{Axiom (E) (Compl{\underline{e}}teness)} For any quantum state over spacetime $\rho_{AE}$ with two arbitrary regions $A$ and $E$ in spacetime, and any quantum channel $\mcal{E}_{B|A}$, the action of state over time function on a subsystem  $\mcal{E}_{B|A}\star \rho_{AE}$ can be defined and has the following properties:
For any completely positive trace non-increasing operation $\mcal{I}_E$ on system $E$,
\begin{align} 
    \mcal{I}_E\qty[\mcal{E}_{B|A}\star \rho_{AE}] &= \mcal{E}_{B|A}\star \mcal{I}_E(\rho_{AE}),\label{eqn:Eind}\\    \Tr_A\qty[\mcal{E}_{B|A}\star \rho_{AE}] &= (\mcal{E}_{B|A}\otimes \id_E) (\rho_{AE}),\label{eqn:prtl}
\end{align}

\textbf{Axiom (P) (Com{\underline{p}}ositionality, \cite{fullwood2023quantum})} A state over time function should be compatible with composition of quantum channels. In other words, for any two quantum channels $\mcal{E}_{B|A}$ and $\mcal{F}_{C|B}$, we have
    \begin{equation} \label{eqn:comp}
        \Tr_B\qty[\mcal{F}_{C|B}\star\qty(\mcal{E}_{B|A}\star \rho_A)]=\qty(\mcal{F}\circ\mcal{E})_{C|A}\star\rho_A.
    \end{equation}

\textbf{Axiom (CC) (Classical Conditionability)} For any quantum channel $\mcal{E}_{B|A}$ such that $\mcal{E}_{B|A}\circ\qty(\sum_i \Ad_{\mds{1}_{A_i}})=\mcal{E}_{B|A}$ where $A=\bigoplus_i A_i$, when $\qty{\lambda_i}$ is a probability distribution, we have
    \begin{equation}
        \mcal{E}_{B|A}\star \qty(\sum_i \lambda_i \pi_{A_i}) = \sum_i \lambda_i \mcal{E}_{B|A_i}\star \pi_{A_i},
    \end{equation}
    where $\mcal{E}_{B|A_i}$ is the limitation of $\mcal{E}_{B|A}$, i.e. $\mcal{E}_{B|A_i}=\mcal{E}_{B|A}\circ\Ad_{\mds{1}_{A_i}}$. \\

\textbf{Axiom (T) (Time reversal symmetry)} A state over time corresponding to the trivial evolution should be symmetric under the time reversal transformation, i.e.
    \begin{equation}
        F_{AB} (\id_{B | A}\star \rho_A)F_{AB}=\id_{B| A}\star \rho_A,
    \end{equation}
    for all $\rho_A\in\mf{S}(A)$.

\end{tcolorbox}

The axioms above require crucial operational properties that one would inherently expect of any state over time function. In the following sections, we will discuss the implications and relations of the axioms given above.

\textit{On linearity.} ---   Eq.(\ref{eqn:prtl}) is the completion of Eq.(\ref{eqn:marg2}) in the sense that we require $\Tr_A\circ \;(\mcal{E}_{B|A}\star \cdot_A)=\mcal{E}_{B|A}(\cdot)$ in the multipartite setting. In the language of Ref. \cite{fullwood2023quantum}, the \textit{bloom-shriek factorization} should be possible. We call this the completeness axiom, following how complete positivity is positivity of a linear map acting on any subsystem of a joint system.

The completeness axiom is very intuitive yet power enough to induce many useful properties. First, one can easily observe that the following properties follow from axiom \textbf{(E)}.
\begin{align} 
    \Tr_{E}\qty[\mcal{E}_{B|A}\star \rho_{AE}] &= \mcal{E}_{B|A}\star \rho_{A},\label{eqn:Trcomm}\\    
    \mcal{E}_{B|A}\star(\rho_A\otimes \sigma_E)
    &=(\mcal{E}_{B|A}\star\rho_A)\otimes \sigma_E.\label{eqn:tens}
\end{align}
Actually, the completeness axiom is equivalent to that the given state over time function is state-linear (proofs in Appendix \ref{app:stlin}).

\begin{prop} \label{prop:stlin}
    If a state over time function $\star$ satisfies \textbf{(E)}, then it is state-linear.
\end{prop}

\begin{corollary} \label{coro:TrEidE}
    For any state over time function $\star$ satisfying axiom \textbf{(E)}, the following holds for any state $\rho_{AE}$ over spacetime,
    \begin{equation} \label{eqn:eqlinex}
        \mcal{E}_{B|A} \star \rho_{AE} = \qty((\mcal{E}_{B|A} \star \;\cdot\;) \otimes \id_E)(\rho_{AE}).
    \end{equation}
    Conversely, every state-linear state over time function satisfies axiom \textbf{(E)} through Eq. (\ref{eqn:eqlinex}).
\end{corollary}

Next, we show that by assuming that a state over time function satisfying axiom \textbf{(E)} is consistent with composition of quantum channels, the state over time function is essentially decided by the corresponding \textit{time-expansion} $\rho_A \mapsto (\id_{A'| A}\star \rho_A)$, a basic function that expands a single-time state $\rho_A$ into a state over two time steps. 

\begin{prop} \label{prop:prlin}
    If a state over time function $\star$ satisfies \textbf{(E)}, then axiom \textbf{(P)} is equivalent to that for any $\mcal{E}_{B|A}\in\mf{C}(A,B)$ and $\rho_A\in\mf{S}(A)$,
    \begin{equation} \label{eqn:Edecomp}
        \mcal{E}_{B|A}\star \rho_A = (\id_A \otimes \mcal{E}_{B|A'}) (\id_{A'| A}\star \rho_A).
    \end{equation}
    Moreover, it implies that $\star$ is process-linear, and the definition of $\star$ can be linearly extended to arbitrary linear maps $\mcal{E}_{B|A}$ that may not be a quantum channel through Eq.(\ref{eqn:Edecomp}). 
\end{prop}
\begin{proof}
    By Eq.(\ref{eqn:comp}), $\Tr_{A'}\qty[\mcal{E}_{B|A'}\star\qty(\id_{A'| A}\star \rho_A)]=\mcal{E}_{B|A}\star\rho_A.$ Furthermore, from Eq.(\ref{eqn:prtl}) it follows that the left-hand side equals to $(\id_A \otimes \mcal{E}_{B|A'}) (\id_{A'| A}\star \rho_A)$. The converse is immediate. For the second part, see Remark \ref{rmk:prlin} in Appendix \ref{app:rmks}.
\end{proof}

In other words, axioms \textbf{(E)} and \textbf{(P)} force the state over time function to be bilinear. Conversely, failure of fulfilling bilinearity leads to the malfunction of state over time function in multipartite settings. This partially answers, in the negative, one of the open problems on the possibility of operationally meaningful state over time functions that are process-nonlinear~\cite{parzygnat2023time}. We remark that not every bilinear state over time function satisfies axiom \textbf{(P)}. (See Remark \ref{rmk:bilin} in Appendix \ref{app:rmks}.)

Although it is a very natural property to expect, not every known state over time function satisfies axiom \textbf{(E)}. As pointed out in Section VII. B. of Ref. \cite{leifer2013towards}, the completeness axiom is not satisfied by the Leifer-Spekkens state over time function defined as $\mcal{E}_{B|A}\star_{ LS}\rho_A:=(\sqrt{\rho}_A\otimes\mds{1}_B)\mscr{D}\qty[\mcal{E}](\sqrt{\rho}_A\otimes\mds{1}_B).$ Considering how widely the Petz recovery map, the Bayesian retrodiction map induced by the Leifer-Spekkens function, is used in quantum thermodynamics \cite{kwon2019fluctuation}, the failure in satisfying axiom \textbf{(E)} poses a huge conceptual problem that we hope to be solved in future works.

\begin{remark}[On compositionality]\label{rmk:comp}
    The compositionality axiom is arguably a simpler and more operational alternative to the associativity axiom \footnote{\begin{equation} \label{eqn:FPassoc}
        \mscr{D}^{-1}\qty[|A| \mcal{F}_{C|B}\star\qty(\frac{\mscr{D}[\mcal{E}_{B|A}]}{|A|})] \star \rho_A=\mcal{F}_{C|B}\star\qty(\mcal{E}_{B|A}\star\rho_A). 
    \end{equation}} given in Ref. \cite{fullwood2022quantum}. Perhaps a simpler form of associativity would be the one given in Section 4 of Ref. \cite{fullwood2022quantum},
    \begin{equation} \label{eqn:assoc}
        (\mcal{F}\star\mcal{E})\star \rho = \mcal{F}\star(\mcal{E}\star \rho),
    \end{equation}
    where the star product of $\mcal{E}$ and $\mcal{F}$ is understood as $(\mcal{F}\star\mcal{E})(\sigma):=\mcal{F}\star \mcal{E}(\sigma)$ for all $\sigma\in\mf{S}(A)$. 
    However, it is not evident if Eq. (\ref{eqn:assoc}) is applicable to state over time functions that may not satisfy Eq. (\ref{eqn:Edecomp}), because it is not immediate if  $\mcal{F}\star\mcal{E}$ is a valid input for the first argument of $\star$, when it is unclear if the definition of $\star$ can be extended to arbitrary linear maps.
    On the other hand, Proposition \ref{prop:prlin} shows that axioms \textbf{(E)} and \textbf{(P)} circumvent this issue, and thus we have $(\mcal{F}_{C|B}\star\mcal{E}_{B|A})\star \rho_A=[(\mcal{F}_{C|B}\star \cdot_B)\circ\; \mcal{E}_{B|A}] \star \rho_A=[(\mcal{F}_{C|B}\star \cdot_B)\circ\; \mcal{E}_{B|A'}] (\id_{A'| A} \star \rho_A)=\mcal{F}_{C|B}\star(\mcal{E}_{B|A}\star \rho_A)$, i.e. Eq. (\ref{eqn:assoc}) follows. Hence, we conclude that the compositionality axiom \textbf{(P)} can safely replace the associativity axiom (\ref{eqn:assoc}) whenever we assume state-linearity.
    We also note that requiring compositionality for any two quantum channels is sufficient for ensuring compositionality for arbitrarily many channels (Appendix \ref{app:comp}).
    
\end{remark}

\textit{On classical limit.} --- We naturally want a given state over time function to be reduced to a classical inference map in the classical setting. The \textit{classical limit axiom} that has been widely employed \cite{horsman2017can, fullwood2022quantum, parzygnat2023time} is given as
    \begin{equation} \label{eqn:class}
        \qty[\mscr{D}[\mcal{E}],\rho_A\otimes\mds{1}_B]=0 \implies \mcal{E}_{B|A}\star \rho_A = \mscr{D}[\mcal{E}](\rho_A\otimes\mds{1}_B),
    \end{equation}
    However, it appears to be stronger than what it aims to achieve. One direct consequence of Eq.(\ref{eqn:class}) is that the state over time for the maximally mixed input state $\pi_A$ is the channel state $\mscr{D}[\mcal{E}]$ up to the normalization factor. this requirement, favouring the \jami state over other alternatives such as the Choi matrix, is a quantum-exclusive feature that does not appear in classical systems. We first explicitly spell it out as an independent axiom.
    \begin{tcolorbox}[breakable, enhanced jigsaw, colback=white, colframe=black]    
        \textbf{Axiom (J) (Jamio{\l}kowski)} For any system $A$ which is possibly a subspace of a larger system, the state over time associated with the maximally mixed state $\pi_{A}:=|A|^{-1}\mds{1}_{A}$ is 
            \begin{equation} \label{eqn:noevent}
                \mcal{E}_{B| A}\star \pi_{A} = \frac{1}{|A|}(\id_A\otimes \mcal{E}_{B|A'})(F_{AA'}).
            \end{equation}
    \end{tcolorbox}

    Indeed, the following result shows that the classical limit axiom is composed of axioms \textbf{(CC)} and \textbf{(J)}. The proof is presented in Appendix \ref{app:J}.
    
    \begin{prop} \label{prop:J}
        Axioms \textbf{(CC)} and \textbf{(J)} are equivalent to the classical limit axiom as represented by Eq.(\ref{eqn:class}).
    \end{prop}

    A question that logically ensues is: Can we weaken the constraint of the classical limit axiom, say, to either axiom \textbf{(CC)} or \textbf{(J)}? In studying their relation, we also investigate yet another axiom that has been often used to impose reduction to classical limit.

    \begin{tcolorbox}[breakable, enhanced jigsaw, colback=white, colframe=black]
        \textbf{Axiom (QC) (Quantum Conditionability)} For every state $\rho \in \mf{S}(A)$, there exists a linear map $\Theta_\rho$ called a \textit{state-rendering function} \cite{hayashi2016quantum,tsang2022generalized, parzygnat2023time} on $\mf{B}(A)$ such that
            \begin{equation} \label{eqn:Qcond}
                \mcal{E}_{B|A}\star \rho_A=(\Theta_\rho\otimes\id_B)\qty(\mcal{E}_{B|A}\star \mds{1}_A ),
            \end{equation}
        for all $\mcal{E}\in\mf{C}(A,B)$ with the property that when $[\rho,M]=0$, we have $\Theta_\rho(M)=\rho M$ for any $M\in \mf{B}(A)$ and $\rho \in \mf{S}(A)$.
    \end{tcolorbox}

Axiom \textbf{(QC)} is employed to enable conditioning on arbitrary input states $\rho$ through $\Theta_\rho$ and the conditional quantum state ${\mcal{E}_{B|A}\star\mds{1}_A}$. The motivation behind it is to form a causally neutral formulation of quantum theory resembling classical probability theory where conditioning is always possible. Therefore, it formally generalizes its classical counterpart; the joint distribution matrix $J_{yx}:=P(X=x,Y=y)$ can be obtained from the conditional distribution matrix $C_{yx}:=P(Y=y|X=x)$ by the right-multiplication with the diagonal matrix $D^p_{xx}:=P(X=x)$, i.e. $J=CD^p$. One can see that $\Theta_\rho$ corresponds to $D^p$, $\mcal{E}_{B|A}\star\mds{1}_A$ to $C$, and $\mcal{E}_{B|A}\star\rho_A$ to $J$. Observe that the matrix $D^p$ is self-adjoint, positive-semidefinite and furthermore $D^p \ket{x} = P(X=x) \ket{x}$. The condition $[\rho,M]=0 \implies \Theta_\rho(M)=\rho M$ in axiom \textbf{(QC)} is a formal generalization of the last property. The following result shows that we need to assume only one of axiom \textbf{(J)} and \textbf{(CC)} when we already imposed state-linearity and compositionality. See Appendix \ref{app:axtoax} for the proof of the following Proposition.

    \begin{prop} \label{prop:ILPII} Assuming axioms \textbf{(E)} and \textbf{(P)}, axioms \textbf{(CC)} and \textbf{(J)} are equivalent and they respectively imply \textbf{(QC)}.
    \end{prop}

Although the motivation behind the requirement about commutator in axiom \textbf{(QC)} was reduction to classical limit in classical theories with commuting algebras, the following result shows that, without axiom \textbf{(T)}, this often employed condition is only formal and not strong enough to induce either of \textbf{(CC)} and \textbf{(J)} under the same condition. This highlights the importance of a proper quantum conditional state. See Appendix \ref{app:PnonLJ} for the proof.

\begin{prop} \label{prop:PnonLJ}
    Axioms \textbf{(E)} and \textbf{(QC)} imply axiom \textbf{(P)} but not \textbf{(CC)} and \textbf{(J)}. However, axioms \textbf{(T)} and \textbf{(QC)} imply axiom \textbf{(CC)} and \textbf{(J)}.
\end{prop}

Axiom \textbf{(QC)} is logically weaker than \textbf{(CC)} or \textbf{(J)}. What if we strengthen it by requiring the state-rendering function $\Theta_\rho$ to be self-adjoint and positive-semidefinite with respect to the Hilbert-Schmidt inner product like its classical counterpart? Let us call these conditions axiom \textbf{(QC+SA)} and \textbf{(QC+PS)} respectively. Note that the latter implies the former. The following Proposition shows that it can narrow down the set of state over time functions, but not to a unique one. The proof can be found in Appendix \ref{app:nonU}. 

\begin{prop}\label{prop:PFuni}  A self-adjoint state-rendering function that is linear in $\rho$ must be of the form 
\begin{align}
    \Theta^\mu_\rho(M)= \mu \rho M + (1-\mu) M \rho
\end{align}
for real number $\mu$. If $\Theta_\rho$ is also positive-semidefinite , then $\mu$ is between 0 and 1.
\end{prop}

Two extreme cases of such functions are known as the \textit{left bloom} $\Theta^L_\rho (M) := M \rho$ and the \textit{right bloom} $\Theta^R_\rho (M) := \rho M$. However, all the convex sums other than the \textit{symmetric bloom}, $\Theta^S_\rho(M):=\frac{1}{2}(\rho M + M \rho)$ yield a state over time that is not Hermitian, and would yield a time-expansion function $\id_{B| A}\star \rho_A$ that is asymmetric under time reversal. We will discuss more about Hermiticity in the next Section. 

We remark that the fact that axiom \textbf{(J)} can be applied to the cases where $A$ is a subspace of a larger Hilbert space was critical. This is physically plausible; for example, a system with an energy cutoff could be considered a system on its own. If we do not accept this and consider a weaker version of axiom \textbf{(J)} called \textbf{(\^{J})} that is only applied to full systems, then the equivalence with \textbf{(CC)} under \textbf{(E)} and \textbf{(P)} breaks down. (See Appendix \ref{app:nonU}.)

\textit{On Hermiticity and time reversal symmetry.} --- We advocate that the arrow of time should not be imposed at the level of state over time function, but instead \textit{emerge} from the correlation of a given state over time. Thus, we claim that a swap between future and past quantum systems $M\otimes N \mapsto N\otimes M$ for all $N\in\mf{B}(A)$ and $M\in \mf{B}(B)$ should be sufficient to describe time reversal because the distinction between future and past is arbitrary. On contrary, in Ref. \cite{parzygnat2023time}, the time reversal map on a bipartite state over time is given as $M\otimes N \mapsto N^\dag \otimes M^\dag$. In Ref. \cite{parzygnat2023time}, it was shown that appending the dagger operation makes certain non-Hermitian state over time functions yield legitimate quantum Bayes' rules, but on the basis of the aforementioned reason, we claim that should instead be a reason to reject non-Hermitian state over time functions. Nevertheless, one might want to impose Hermiticity to states over time directly as it was done in the previous works \cite{horsman2017can, fullwood2022quantum, parzygnat2023time, fullwood2023quantum}.
     
    \begin{tcolorbox}[breakable, enhanced jigsaw, colback=white, colframe=black]
        \textbf{Axiom (H) (Hermiticity)} For any quantum channel $\mcal{E}_{B|A}$ and any quantum state $\rho_A$, the state over time $\mcal{E}_{B|A}\star\rho_A$ must be a Hermitian operator.
    \end{tcolorbox}
    As Remark \ref{rmk:comp} and Propositions \ref{prop:prlin}-\ref{prop:J} show, axioms \textbf{(H)}, \textbf{(E)}, \textbf{(CC)}, \textbf{(P)} and \textbf{(J)} imply all the axioms considered before in Ref.\cite{horsman2017can, fullwood2022quantum, parzygnat2023time, fullwood2023quantum}, \textit{Hermiticity, bilinearity, preservation of classical limit} and \textit{associativity} (except for positivity, see Appendix \ref{app:rmks}.) However, it turns out that axiom \textbf{(H)} is not as strong as axiom \textbf{(T)} because it cannot induce a unique state over time function. (See Appendix \ref{app:insuff} for the proof.) Proposition \ref{prop:PFuni} tells us that axiom \textbf{(H)} requires a stronger version of axiom \textbf{(QC)}, axiom \textbf{(QC+SA)}, to uniquely characterize a state over time function. We can summarize the results of this work as follows.

\begin{theorem} \label{thm:equiv}
    The following combinations of axioms are all equivalent and satisfied only by the FP function:
    \begin{center}
        \begin{tabular}{ccc}
          \textbf{(E)+(P)+(CC)+(T)} & or & \textbf{(E)+(P)+(J)+(T)} \\
           or \; \textbf{(E)+(QC)+(T)} & or & \textbf{(E)+(QC+SA)+(H)}.
        \end{tabular}
    \end{center}
\end{theorem}

\textit{Application to acausal regions.}---   Despite its name, state over time functions can also be used to find conditional quantum state associated with a given bipartite quantum state over \textit{space} \cite{leifer2013towards}. For any $\rho_{AB}\in\mf{S}(AB)$, if the spectral decomposition of $\rho_A$ is given as $\rho_A=\sum_i \lambda_i \dyad{\lambda_i}$, then the inverse of the corresponding state-rendering function is given as
    \begin{equation}
        (\Theta^{S}_\rho)^{-1}(\sigma)=\sum_{ij} \frac{2}{\lambda_i+\lambda_j} \bra{\lambda_i}\sigma\ket{\lambda_j} \dyad{\lambda_i}{\lambda_j}.
    \end{equation}
    Using this expression, one can calculate the conditional quantum state $\rho_{A|B}$ corresponding to arbitrary bipartite state $\rho_{AB}$, i.e. $\rho_{B|A}=((\Theta^{S}_\rho)^{-1}\otimes \id_B)(\rho_{AB})$ and moreover the corresponding \textit{belief propagation map} $\mcal{B}_\rho$ \cite{leifer2013towards} via $\mcal{B}^\rho_{B|A}(\sigma_A):=\Tr_A[(\sigma_A\otimes\mds{1}_B)\rho_{B|A}],$ and they are related with each other through $\rho_{A|B}=\mscr{D}[\mcal{B}^\rho_{B|A}]$.
    Notably, one does not have to specify if $\rho_{AB}$ is a state over time or over space to conduct this calculation. This shows that the FP function is not only a state over time, but also a state over \textit{space} function, therefore Theorem \ref{thm:PF} characterizes the FP function as the unique representation of quantum states over spacetime.

\textit{Conclusions.}--- (1) We showed that the axioms introduced in literature do not yield a unique state over time function and (2) introduced a set of operationally motivated axioms for state over time with focus on its application in multipartite settings that is stronger than the one given in Ref. \cite{fullwood2022quantum}. (3) We analyzed the (in)equivalence relations between alternative sets of axioms, and (4) characterized the Fullwood-Parzygnat state over time function as the unique state over time function. An interesting aspect of our proof is that the mathematical technique \cite{heunen2013matrix} once used to prove the no-go result for state over time \cite{horsman2017can} is used to prove the uniqueness of the state over time function here.


With the establishing of uniqueness for the state over time function, we establish a landmark result on the efforts to construct a causally neutral framework of quantum theory. Further attempts can now be focused on discovering the various applications of this formalism as the usefulness of the FP function is already demonstrated by recent applications \cite{fullwood2023dynamical, fullwood2023quantum}.

\acknowledgements
Upon completion of the manuscript, we became aware of an independent work of Parzygnat, Fullwood, Buscemi and Chiribella that characterizes the FP function from a slightly different set of axioms \cite{parzygnat2023virtual}. SHL thanks A. J. Parzygnat and J. Fullwood for helpful discussions. This work was supported by the start-up grant of the Nanyang Assistant Professorship awarded to Nelly H.Y. Ng of Nanyang Technological University, Singapore.


\appendix
\section{Extension of compositionality} \label{app:comp}

\begin{prop}
    For any state over time function $\star$ satisfying axiom \textbf{(E)} and any compatible channels $\mcal{E,F}$ and $\mcal{G}$, we have
    \begin{equation} \label{eqn:2comp}
    \Tr_{BC}\qty[\mcal{G}_{D|C}\star \qty(\mcal{F}_{C|B}\star\qty(\mcal{E}_{B|A}\star \rho_A))]=\qty(\mcal{G}\circ\mcal{F}\circ\mcal{E})_{D|A}\star\rho_A.
\end{equation}
\end{prop}

\begin{proof}
    We first claim that for any state over spacetime $\eta_{AEF}$,
    \begin{equation}
    \Tr_E\qty[\mcal{E}_{B|A}\star \eta_{AEF}] := \mcal{E}_{B|A}\star \eta_{AF},
\end{equation}
    which follows from simply combining two systems $AF$ into $A$ and replacing $\mcal{E}_{B|A}$ with $\mcal{E}_{B|A}\otimes \Tr_F$ in Eq. (\ref{eqn:Trcomm}). Let $\sigma_{ABC}:=\mcal{F}_{C|B}\star\qty(\mcal{E}_{B|A}\star \rho_A)$. Then, the left hand side of Eq. (\ref{eqn:2comp}) gives
    \begin{equation}
        \Tr_C\qty[\mcal{G}_{D|C}\star \Tr_B\sigma_{ABC}]
    \end{equation}
    whereas $\Tr_B\sigma_{ABC}=\qty(\mcal{F}_{C|B}\circ \mcal{E}_{B|A})\star \rho_A$ by the compositionality axiom. Therefore, by invoking compositionality again, we obtain the right hand side of Eq. (\ref{eqn:2comp}).
\end{proof}

\section{Proof of Proposition \ref{prop:stlin} and Corollary \ref{coro:TrEidE}} \label{app:stlin}
We first give the proof of Proposition \ref{prop:stlin}.
\begin{proof}
    Let $E$ be a two level system. For any two states $\rho^{0,1}_A$ on $A$ and arbitrary $0\leq \lambda \leq 1$, consider the following quantum-classical state $\rho_{AE}$,
    \begin{equation}
        \rho_{AE}=\lambda \rho^0_A\otimes\dyad{0}_E + (1-\lambda)\rho^1_A\otimes \dyad{1}_E.
    \end{equation}
    For any $\mcal{E}_{B|A}\in\mf{C}(A,B)$, we can define a state over time
    \begin{equation}
        \tau_{BAE}=\mcal{E}_{B|A}\star\rho_{AE}.
    \end{equation}
    We now observe that $\rho_{AE}$ is invariant under the action of the dephasing channel on $E$ given as $\mcal{D}_E=\Ad_{\dyad{0}_E}+\Ad_{\dyad{1}_E}$. From Eq. (\ref{eqn:Eind}), it follows that $\tau_{BAE}$ is also invariant under $\mcal{D}_E$. Hence it is also of the following form.
    \begin{equation} \label{eqn:taubae}
        \tau_{BAE}=c_0 \tau^0_{BA}\otimes \dyad{0}_E + c_1 \tau^1_{BA}\otimes \dyad{1}_E,
    \end{equation}
    where $\Tr\tau^{0,1}_{BA}=1$ are normalized so that $c_0+c_1=1$. Apply $\Ad_{\dyad{0}_E}$ to both hand sides of \ref{eqn:taubae} to get 
    \begin{equation}
        \mcal{E}_{B|A}\star(\lambda \rho^0_A \otimes \dyad{0}_E)= c_0 \tau^0_{BA} \otimes \dyad{0}_E,
    \end{equation}
    by Eq. (\ref{eqn:Eind}), again. After tracing out $E$ from both sides, we get $\tau^0_{BA}=\mcal{E}_{B|A}\star\rho^0_A$ and $c_0=\lambda$. Similarly we can get $\tau^1_{BA}=\mcal{E}_{B|A}\star\rho^1_A$ and $c_1=1-\lambda$. Therefore, by taking $\Tr_E$ of Eq. (\ref{eqn:taubae}), we get
    \begin{equation}
        \mcal{E}_{B|A}\star\qty(\lambda\rho^0_A+(1-\lambda)\rho^1_A)=\lambda \mcal{E}_{B|A}\star\rho^0_A+(1-\lambda)\mcal{E}_{B|A}\star\rho^1_A.
    \end{equation}
    It follows that the star product $\star$ is state-linear.
\end{proof}
Next, we present the proof of Corollary \ref{coro:TrEidE}.
\begin{proof}
    Every operator $\rho_{AE}\in\mf{B}(AB)$ can be expressed as a linear combination of product states of the form of $\rho_A\otimes \sigma_E$. By Proposition \ref{prop:stlin} or assumption, the star product is state-linear, hence, through Eq. (\ref{eqn:tens}) we get the wanted result.
\end{proof}

\section{Proof of Proposition \ref{prop:J}} \label{app:J}

    \begin{proof}
        Assume axioms \textbf{(CC)} and \textbf{(J)}. Let $\qty{A_i}$ be eigenspaces corresponding to distinct eigenvalues $\qty{\lambda_i}$ of $\rho$ so that $\rho=\sum_i \lambda_i \mds{1}_{A_i}$. The condition $[\mscr{D}[\mcal{E}_{B|A}],\rho_A\otimes \mds{1}_B]=0$ is equivalent to
        \begin{equation}
            \sum_i (\mds{1}_{A_i}\otimes \mds{1}_B)\mscr{D}[\mcal{E}_{B|A}](\mds{1}_{A_i}\otimes\mds{1}_B)=\mscr{D}[\mcal{E}].
        \end{equation}
        It is in turn equivalent to $\sum_i\mcal{E}_{B|A}\circ\Ad_{\mds{1}_{A_i}}=\mcal{E}_{B|A}$. Therefore, by axiom \textbf{(CC)},
        \begin{equation}
            \mcal{E}_{B|A}\star\rho_A=\sum_i \lambda_i \mcal{E}_{B|A_i}\star \mds{1}_{A_i}.
        \end{equation}
         By axiom \textbf{(J)}, $\mcal{E}_{B|A_i}\star\pi_{A_i}=|A_i|^{-1}(\id_A\otimes \mcal{E}_{B|A_i'})(F_{A_iA'_i})$. Hence, we have
        \begin{equation}
            \mcal{E}_{B|A}\star \rho_A =\sum_i \lambda_i (\id_{A}\otimes \mcal{E}_{B|A_i'})(F_{A_iA'_i}).
        \end{equation}
        Observe that $(\id_{A}\otimes \mcal{E}_{B|A_i'})(F_{A_iA'_i})=(\Ad_{\mds{1}_{A_i}}\otimes\mcal{E}_{B|A'})(F_{AA'})$. So it follows that
        \begin{equation}
            \mcal{E}_{B|A}\star \rho_A=\sum_i \lambda_i (\Ad_{\mds{1}_{A_i}}\otimes \id_{B}) (\mscr{D}[\mcal{E}_{B|A}]).
        \end{equation}
        Note that the right hand side is equal to $\mscr{D}[\mcal{E}_{B|A}](\rho_A\otimes \mds{1}_B)$.

        Conversely, assume that $[\mscr{D}[\mcal{E}_{B|A}],\rho_A\otimes \mds{1}_B]=0$ implies $\mcal{E}_{B|A}\star\rho_A=\mscr{D}[\mcal{E}_{B|A}](\rho_A\otimes\mds{1}_B)$. Axiom \textbf{(J)} immediately follows considering the case of $\rho_A=\pi_A$. As we mentioned before, $[\mscr{D}[\mcal{E}_{B|A}],\rho_A\otimes \mds{1}_B]=0$ is equivalent to $\sum_i\mcal{E}_{B|A}\circ\Ad_{\mds{1}_{A_i}}=\mcal{E}_{B|A}$, when $\rho=\sum_i \lambda_i  \mds{1}_{A_i}$ with distinct $\lambda_i$'s. In this case, $\mscr{D}[\mcal{E}_{B|A}](\rho_A\otimes\mds{1}_B)=\sum_i \lambda_i \mscr{D}[\mcal{E}_{B|A_i}]$, but $\mscr{D}[\mcal{E}_{B|A_i}]=\mcal{E}_{B|A_i}\star \mds{1}_{A_i}$ by axiom \textbf{(J)}, hence axiom \textbf{(CC)} also follows.
    \end{proof}

\section{Further remarks on axioms} \label{app:rmks}
\begin{remark} [On process-linearity] \label{rmk:prlin}
    There is one mathematical subtlety about state over time functions satisfying Eq. (\ref{eqn:Edecomp}) when one tries to extend the domain of the first argument from the set of quantum channels to the set of all linear maps from $\mf{B}(A)$ to $\mf{B}(B)$. It is that the extension is not uniquely decided because the span of the set of quantum channels $\mf{C}(A,B)$ does not contain the linear maps of the form
    \begin{equation}
        \mcal{L}_{\sigma}(M):=\mds{1}_B \Tr[\sigma M],
    \end{equation}
    for all $M\in\mf{B}(A)$ and $\sigma\in\mf{B}(B)$ such that $\Tr[\sigma]=0$. This opens up the possibility that a given star product $\star$ has two different time-expansion functions for the span of quantum channels and the linear maps $\mcal{L}_\sigma$ given above. For example, one can define a state over time function $\mcal{E}_{B|A}\star\rho_A$ as the linear extension of the map given as
    \begin{equation}
        (\id_A\otimes\mcal{E}_{B|A'})((\rho_A\otimes\mds{1}_B) F_{AA'}), 
    \end{equation}
    for $\mcal{E}_{B|A}\in\langle\mf{C}(A,B)\rangle$ (the span of $\mf{C}(A,B)$) and
    \begin{equation}
        (\id_A\otimes \mcal{L}_\sigma) (F_{AA'}(\rho_A\otimes\mds{1}_B)), 
    \end{equation}
    for $\mcal{E}_{B|A}=\mcal{L}_\sigma$ for some traceless $\sigma\in\mf{B}(B)$. Then, $\star$ is a bilinear state over time function, but Eq. (\ref{eqn:Edecomp}) does not hold.
    Nevertheless, there always \textit{exists} an extension of $\star$ defined as 
    \begin{equation}
        \mcal{E}_{B|A}\star\rho_A:=(\id_A\otimes \mcal{E}_{B|A'})(\id_{A'|A}\star\rho_A),
    \end{equation}
    for all $\mcal{E}_{B|A}\in\mf{B}(\mf{B}(A),\mf{B}(B))$.
\end{remark}

\begin{remark}[On bilinearity] \label{rmk:bilin}
    Not every bilinear state over time function is decomposable in the form of Eq. (\ref{eqn:Edecomp}). For example, for such a decomposable state over time function $\star,$ we define a new function $\star'$
    \begin{equation}
        \mcal{E}_{B|A}\star'\rho_A:=\mcal{E}_{B|A}\star\rho_A + \Tr[\mcal{E}(\rho)]\Xi_{AB},
    \end{equation}
    for some $\Xi\in\mf{B}(AB)$ such that both $\Tr_A\Xi_{AB}$ and $\Tr_B\Xi_{AB}$ vanish. One can easily check that for any quantum channel $\mcal{E}_{B|A}$ and any quantum state $\rho_A$ $\Tr[\mcal{E}(\rho)]=1$ so that the second term is constant. If there is a quantum channel $\mcal{F}_{B|A}$ such that $(\id_A\otimes \mcal{F}_{B|A})(\Xi_{AB})\neq\Xi_{AB}$, then it violates the assumption that $\star'$ is decomposable.
\end{remark}

\begin{remark}[On linear extensions] \label{rmk:linear}
    Although we did not elaborate in the main text, Proposition \ref{prop:stlin} only shows the convex-linearity in state of the star product, not the full linearity. However, one can recover the full linearity as follows. First, recall that any Hermitian operator $X\in\mf{H}(A)$ can be expressed as $X=\Tr[X]\mds{1}_A/|A| +\lambda (\rho-\sigma)$ for some $\rho,\sigma \in \mf{S}(A)$ and $\lambda\geq 0$. For any convex-linear function on $\mf{S}(A)$, we can uniquely extend its domain to $\mf{H}(A)$ by letting
    \begin{equation} \label{eqn:Hdecomp}
        f(X):=\Tr[X]f\qty(\frac{\mds{1}_A}{|A|})+\lambda \qty(f(\rho)-f(\sigma)).
    \end{equation}
    This extension is well-defined because whenever $\lambda(\rho-\sigma)=\lambda'(\rho'-\sigma')$, i.e. $\lambda\rho+\lambda'\sigma' = \lambda'\rho'+\lambda\sigma$ with some $\lambda'\geq 0$ and $\rho',\sigma'\in\mf{S}(A)$, by the convex-linearity of $f$ on $\mf{S}(A)$, (without loss of generality we assume $
    \lambda+\lambda' >0$) we have
    \begin{equation}
        f\qty(\frac{\lambda}{\lambda+\lambda'}\rho + \frac{\lambda'}{\lambda+\lambda'}\sigma')=\frac{\lambda}{\lambda+\lambda'}f(\rho)+\frac{\lambda'}{\lambda+\lambda'}f(\sigma'),
    \end{equation}
    and similarly
    \begin{equation}
        f\qty(\frac{\lambda}{\lambda+\lambda'}\sigma + \frac{\lambda'}{\lambda+\lambda'}\rho')=\frac{\lambda}{\lambda+\lambda'}f(\sigma)+\frac{\lambda'}{\lambda+\lambda'}f(\rho').
    \end{equation}
    Therefore, we have $\lambda f(\rho) + \lambda' f(\sigma') = \lambda f(\sigma) + \lambda' f(\rho'),$ i.e. $\lambda(f(\rho)-f(\sigma))=\lambda'(f(\rho')-f(\sigma'))$. Thus $f$ can be uniquely extended to a linear function on $\mf{H}(A)$; uniqueness can be seen from the fact that Eq. (\ref{eqn:Hdecomp}) is satisfied by any linear function on $\mf{H}(A)$ so that the value of $f$ is completely determined if its values on quantum states are fixed. Furthermore, by observing that any non-Hermitian operator $X$ can be decomposed into the Hermitian and the anti-Hermitian parts, i.e.
    \begin{equation}
        X=\frac{X+X^\dag}{2} + i \, \frac{X-X^\dag}{2i},
    \end{equation}
    one can uniquely extend $f$ to a linear function on $\mf{B}(X)$ by letting
    \begin{equation}
        f(X):=f\qty(\frac{X+X^\dag}{2})+if\qty(\frac{X-X^\dag}{2i}).
    \end{equation}
\end{remark}

\begin{remark}[On lack of positivity] \label{rmk:nonP}
    One may be tempted to reject our axioms because, throughout the axioms listed in the main text, we did not require states over time to be positive, or even block-positive \cite{johnston2013duality}. However, we claim that this lack of positivity is not a flaw; conventional quantum states are required to be positive because we want them to produce non-negative probability for arbitrary measurement outcome, i.e. $\Tr[M\rho]\geq 0$ for arbitrary POVM element $M$. 
    
    It is, nevertheless, a priori not obvious why arbitrary multipartite measurement should be allowed on quantum states over time. By employing the framework of state over time, one chooses to view the quantum dynamics in a static picture where every history of quantum operation is decided. Crucially, quantum measurement cannot be implemented non-invasively; quantum measurement itself is an operation, and one cannot expect to obtain physical measurement statistics after intervening the predetermined history of quantum systems. It is akin to the phase space in classical mechanics or the Minkowski diagram in relativity \cite{fullwood2023quantum}. Movement of objects are already decided in those pictures and altering the state at one moment affects the whole history.  
\end{remark}

\begin{remark} [On classical conditionability] \label{rmk:appcl}
    In this work, we did not introduce the concept of partially classical systems \cite{lie2023faithfulness}, i.e. quantum systems whose associated $C^*$-algebra is the direct sum of matrix algebras sometimes called multimatrix algebras \cite{fullwood2022quantum, parzygnat2023time, fullwood2023quantum}. Using a multimatrix algebra is equivalent to imposing a superselection rule to the system so that no coherence between certain subspaces of the associated Hilbert space is allowed. We avoided this concept to simplify the mathematical proofs, because a multimatrix algebra can always be considered a subalgebra of a larger full matrix algebra, and also because one can enforce partial classicality on the level of quantum channels acting on them as we did in axiom \textbf{(CC)}. It means that, when a quantum channel $\mcal{E}_{B|A}$ acts on system $A$, $A$ acts like a partially classical system with respect to the decomposition $A=\bigoplus_i A_i$ if
    \begin{equation} \label{eqn:PC}
        \mcal{E}_{B|A}\circ\qty(\sum_i \Ad_{\mds{1}_{A_i}})=\mcal{E}_{B|A}.
    \end{equation}
    and similarly, every channel $\mcal{F}_{A|C}$ should satisfy
    \begin{equation} \label{eqn:PC2}
        \qty(\sum_i \Ad_{\mds{1}_{A_i}})\circ \mcal{F}_{A|C}=\mcal{F}_{A|C}.
    \end{equation}
    In other words, applying the dephasing channel with respect to the decomposition $A=\bigoplus_i A_i$ before $\mcal{E}_{B|A}$ or after $\mcal{F}_{A|C}$ does not alter the action of them. Declaring $A$ is inherently partially classical with respect to $A=\bigoplus_i A_i$ amounts to following a convention in which every channel acting on $A$ satisfies Eq. (\ref{eqn:PC}) and (\ref{eqn:PC2}) and every quantum state on $A$ is block-diagonal with respect to the decomposition $A=\bigoplus_i A_i$.
\end{remark}

\section{Proof of Proposition \ref{prop:ILPII}} \label{app:axtoax}
We first need the following form of axiom \textbf{(CC)}.
\begin{corollary} \label{coro:limit}
    Assuming axioms \textbf{(E)} and \textbf{(P)}, axiom \textbf{(CC)} is equivalent to that for any orthogonal subspaces $A_i$ and $A_j$ of $A$, we have
    \begin{equation}
        (\id_A\otimes \Ad_{\mds{1}_{A_j'}})(\id_{A'|A}\star\pi_{A_i})=0.
    \end{equation}
\end{corollary}
\begin{proof}
    Consider the dephasing channel $\mcal{D}_A=\Ad_{\mds{1}_{A_i}}+\Ad_{\mds{1}_{A_i}^\perp}$ on $A$ where $\mds{1}_{A_i}^\perp:=\mds{1}_A-\mds{1}_{A_i}$. We will denote $\mcal{D}_{A'|A}=\id_{A'|A}\circ\mcal{D}_A$. From axiom \textbf{(CC)} it follows that
    \begin{equation}
        \mcal{D}_{A'|A}\star\pi_{A_i}=(\id_{A'|A}\circ\Ad_{\mds{1}_{A_i}})\star\pi_{A_i}.
    \end{equation}
    By Proposition \ref{prop:prlin}, it follows that
    \begin{equation}
        \qty(\id_{A'|A}\circ\Ad_{\mds{1}_{A_i}^\perp})\star\pi_{A_i}=0.
    \end{equation}
    Note that $\id_{A'|A}\circ \Ad_{\mds{1}_{A_i}^\perp}=\Ad_{\mds{1}_{A_i}^{\perp '}}\circ\id_{A'|A}$ and  $\Ad_{\mds{1}_{A_j}}\circ\Ad_{\mds{1}_{A_i}^\perp}=\Ad_{\mds{1}_{A_j}}$. Again, from Proposition \ref{prop:prlin}, it follows that
    \begin{equation}
        (\id_A\otimes \Ad_{\mds{1}_{A_j'}})(\id_{A'|A}\star\pi_{A_i})=0.
    \end{equation}
\end{proof}

Now we first show that axioms \textbf{(E)+(P)+(CC)} imply \textbf{(J)} and \textbf{(QC)}.

\begin{proof}
    Note that there exists a one-to-one correspondence \cite{jamiolkowski1972linear} between the time-expansion $\id_{A'|A}\star\rho_A$ and a linear map $\Theta_\rho$ on $\mf{B}(A)$  through,
        \begin{equation} \label{eqn:F1}
            \id_{A'|A}\star\rho_A=(\Theta_\rho\otimes\id_{A'})(F_{AA'}).
        \end{equation}
    Note that this $\Theta_\rho$ need not be the state-rendering function in axiom \textbf{(QC)} at this point, although it will turn out to be so eventually. It immediately follows from axiom \textbf{(E)} through Proposition \ref{prop:stlin} that $\Theta_\rho$ is linear in $\rho$. By Eq. (\ref{eqn:marg2}), after taking partial trace on $A'$ in Eq. (\ref{eqn:F1}), we have $\rho_A=\Theta_\rho(\mds{1}_A)$ for all $\rho\in\mf{B}(A)$. Next, from Corollary \ref{coro:limit}, it follows that whenever $A=\bigoplus_i A_i$ and $i\neq j$,
    \begin{equation}
        \Theta_{\mds{1}_{A_i}}(\mds{1}_{A_j})=0.
    \end{equation}
    Since $\Theta_{\mds{1}_{A_i}}(\mds{1}_A)=\mds{1}_{A_i}$, it implies that
    \begin{equation}
        \mds{1}_{A_i}=\Theta_{\mds{1}_{A_i}}(\mds{1}_A)=\Theta_{\mds{1}_{A_i}}\qty(\sum_i \mds{1}_{A_i})=\Theta_{\mds{1}_{A_i}}(\mds{1}_{A_i})
    \end{equation}
    In summary, we have
    \begin{equation} \label{eqn:Eorth}
        \Theta_{\mds{1}_{A_i}}(\mds{1}_{A_j})=\delta_{ij}\mds{1}_{A_i}.
    \end{equation}
    For any subspace $A_1$ of $A$ and an arbitrary pure state $\ket{\psi}\in A_1$, we have $\mds{1}_A=\dyad{\psi}+(\mds{1}_{A_1}-\dyad{\psi})+(\mds{1}_A-\mds{1}_{A_1})$ with every term orthogonal to each other. So, it follows from Eq. (\ref{eqn:Eorth}) and
    the linearity of $\Theta_\rho$ in $\rho$ that
    \begin{equation} \label{eqn:Fid}
        \Theta_{\mds{1}_{A_1}}(\dyad{\psi})=\Theta_{\dyad{\psi}}(\dyad{\psi})=\dyad{\psi}.
    \end{equation}
    Since $\ket{\psi}$ was an arbitrary state in $A_1$, it immediately follows that $\Theta_{\mds{1}_{A_1}}$ is the identity map on $\mf{B}(A_1)$. For the special case of $A_1=A$, by Eq. (\ref{eqn:F1}), we can conclude that $\id_{A'|A}\star\mds{1}_A=F_{AA'}$, i.e. axiom \textbf{(J)}.

    Next, for any $\rho\in\mf{S}(A)$, let $\rho=\sum_i \lambda_i \mds{1}_{A_i}$ be its spectral decomposition and $A=\bigoplus_i A_i$ be the eigenspace decomposition of $A$. Note that for any $M\in \mf{B}(A)$, $[\rho,M]=0$ is equivalent to $M$ being block-diagonal with respect to $\bigoplus_i A_i$, i.e. $M=\sum_i \mds{1}_{A_i}M\mds{1}_{A_i}$. Because of Eq. (\ref{eqn:Fid}), we can conclude that $\Theta_{\mds{1}_{A_i}}(\mds{1}_{A_i}M\mds{1}_{A_i})=\mds{1}_{A_i}M\mds{1}_{A_i}$. By noting that $\Theta_\rho=\sum_i \lambda_i \Theta_{\mds{1}_{A_i}}$ from the linearity of $\Theta_\rho$ in $\rho$ and that $\rho M =\sum_i \lambda_i \mds{1}_{A_i} M \mds{1}_{A_i}$ if $[\rho,M]=0$, it leads to the conclusion that
    \begin{equation}
        \Theta_\rho(M)=\rho M,
    \end{equation}
    whenever $[\rho,M]=0$. Therefore $\Theta_\rho$ is indeed the state-rendering function in axiom \textbf{(QC)}.
\end{proof}

Next, conversely we show that axioms \textbf{(E)+(P)+(J)} imply \textbf{(CC)} and \textbf{(QC)}.

\begin{proof}
    We start from the correspondence Eq. (\ref{eqn:F1}). We first observe that in terms of $\Theta_\rho$ in Eq. (\ref{eqn:F1}), axiom \textbf{(CC)} is equivalent to
    \begin{equation}
        \sum_{i,j}\lambda_i \Theta_{\mds{1}_{A_i}}\circ \Ad_{\mds{1}_{A_j}}=\sum_i \lambda_i \Theta_{\mds{1}_{A_i}}\circ \Ad_{\mds{1}_{A_i}},
    \end{equation}
    when $A=\bigoplus_i A_i$ and $\rho=\sum_i \lambda_i \mds{1}_{A_i}$. On the other hand, Axiom \textbf{(J)} implies the following in terms of $\Theta_\rho$
    \begin{equation}
        \Theta_{\mds{1}_{A_i}}\circ\Ad_{\mds{1}_{A_i}}=\Ad_{\mds{1}_{A_i}},
    \end{equation}
    after we identify the identity map $\id_{A_i}$ on $\mf{B}(A_i)$ with $\Ad_{\mds{1}_{A_i}}$. Consider the case of $A=A_1\oplus A_2$. For an arbitrary $M=M_1+M_2$ with $M_i\in \mf{B}(A_i)$ for $i=1,2$,
    \begin{equation}
        \Theta_{\mds{1}_A}(M)=\sum_{i,j=1}^2 \Theta_{\mds{1}_{A_i}}(M_j).
    \end{equation}
    Note that since $\Theta_{\mds{1}_A}=\id_A$, the left hand side is $M=M_1+M_2$. Because $\Theta_{\mds{1}_{A_i}}(M_i)=M_i$ for $i=1,2$, we have
    \begin{equation}
        \Theta_{\mds{1}_{A_1}}(M_2)+\Theta_{\mds{1}_{A_2}}(M_1)=0.
    \end{equation}
    Since the choice of $M_i\in\mf{B}(A_i)$ for $i=1,2$ was arbitrary, by setting $M_1=0$, we can conclude that
    \begin{equation}
        \Theta_{\mds{1}_{A_1}}\circ \Ad_{\mds{1}_{A_2}}=0.
    \end{equation}
    In the general case of $A=\bigoplus_i A_i$ with more than two direct summands, we have the following by noting that, for all $j$ that is different from $i$, $\Ad_{\sum_{j(\neq i)} \mds{1}_{A_j}}\circ\Ad_{\mds{1}_{A_j}}=\Ad_{\mds{1}_{A_j}}$,
    \begin{equation}
        \Theta_{\mds{1}_{A_i}}\circ \Ad_{\mds{1}_{A_j}}=\delta_{ij}\Ad_{\mds{1}_{A_i}}.
    \end{equation}
    It proves the wanted result, axiom \textbf{(CC)}. Because of the previous proof, we now have axioms \textbf{(E)+(P)+(CC)} in addition to \textbf{(J)}, axiom \textbf{(QC)} follows, too.
\end{proof}

\section{Proof of Proposition \ref{prop:PnonLJ}} \label{app:PnonLJ}
\begin{proof}
    The following Lemma from Ref. \cite{heunen2013matrix} is crucial for this proof and the main result of this work.
    \begin{lemma} [Proposition 2, \cite{heunen2013matrix}] \label{lem:P2} If a bilnear binary operation $X\odot Y$ defined for $X,Y\in\mf{B}(A)$ satisfies $X \odot \mds{1}_A = X$ and $X\odot Y =0$ whenever $X$ and $Y$ are mutually orthogonal projectors for all $X$ and $Y$ in $\mf{B}(A)$ then it has the following form
    \begin{equation} \label{eqn:XYg}
        X \odot Y = XY + g([X,Y])
    \end{equation}
    for some linear map $g$ on $\mf{B}(A)$.
    \end{lemma}

    First of all, we observe that the state-rendering function $\Theta_\rho(M)$ of a state over time function satisfying axioms \textbf{(E)}, \textbf{(QC)} is a binary operation of $\rho$ and $M$ satisfying all the conditions in Lemma \ref{lem:P2}.
    
    For any state over time function $\star$ that satisfies axioms \textbf{(E)} and \textbf{(QC)}, we let $\Phi^{\mcal{E}}_{B|A}$ be the linear map from $\mf{B}(A)$ to $\mf{B}(B)$ corresponding to $\mcal{E}_{B|A}$ through
    \begin{equation}
        \mcal{E}_{B|A}\star\mds{1}_A = (\id_A \otimes \Phi^{\mcal{E}}_{B|A'})(F_{AA'}).
    \end{equation}
    As a special case, we let $\Phi_A:=\Phi^{\id_A}_{A|A}$. If we let $\Psi(\rho_A):=\Theta^\ddag_\rho(\mds{1}_A)$ for any $\rho_A\in\mf{B}(A)$ where $\mcal{L}^\ddag:=\dag\circ\mcal{L}^\dag\circ\dag$ for any linear map defined on $\mf{B}(A)$, by taking partial trace on $A$ in
    \begin{equation} \label{eqn:F3}
        \mcal{E}_{B|A}\star \rho_A = (\Theta_\rho \otimes \id_B)(\mcal{E}_{B|A}\star \mds{1}_A),
    \end{equation}
    we get
    \begin{equation}
        \mcal{E}_{B|A} = \Phi^\mcal{E}_{B|A} \circ \Psi.
    \end{equation}
    Especially, when $A=B$ and $\mcal{E}_{B|A}=\id_A$, we have $\id_A=\Phi_A\circ \Psi_A$, and it follows that $\Psi_A=\Phi^{-1}_A$. Thus,
    \begin{equation}
        \Phi^\mcal{E}_{B|A}=\mcal{E}_{B|A}\circ\Phi_A,
    \end{equation}
    which is equivalent to
    \begin{equation}
        \mcal{E}_{B|A}\star \mds{1}_A = (\id_A \otimes \mcal{E}_{B|A'})(\id_{A'|A}\star \mds{1}_A)
    \end{equation}
    When combined with (\ref{eqn:F3}), we have axiom \textbf{(P)}.

    Next, assuming axioms \textbf{(E)} and \textbf{(P)}, we observe that axiom \textbf{(CC)} is equivalent to the following in terms of the state-rendering function $\Theta_\rho$:
    \begin{equation} \label{eqn:axPinF}
        \Theta_{\mds{1}_{A_i}}\circ (\Psi^{-1})^\ddag \circ \Ad_{\mds{1}_{A_j}}=0
    \end{equation}
    whenever $A=\bigoplus_i A_i$ and $i\neq j$.
    
    consider the case in which $g=r\Ad_{\dyad{\eta}}$ for some fixed pure state $\ket{\eta}\in A$ with some $0<r<1$. In this case, we have
    \begin{equation}
        \Psi(\rho)=\rho + r[\dyad{\eta},\rho] = (\mds{1}_A + r \dyad{\eta})\rho - r \rho \dyad{\eta}.
    \end{equation}
    Since the spectrum of $\mds{1}_A + r \dyad{\eta}$ is $\qty{1,1+r}$ and that of $r\dyad{\eta}$ is $\qty{r}$, they are lower or upper bounded by $(1+r)/2$. By the Bhatia-Davis-McIntosh theorem \textbf{(Cite here)},
    \begin{equation}
        \Psi^{-1}(\rho)=\qty(\frac{1}{1-r}\dyad{\eta}^\perp + \dyad{\eta})\rho\dyad{\eta},
    \end{equation}
    where $\dyad{\eta}^\perp:=\mds{1}_A-\dyad{\eta}$. Thus,
    \begin{equation}
        (\Psi^{-1})^\ddag(\rho)=\dyad{\eta}\rho\qty(\frac{1}{1-r}\dyad{\eta}^\perp + \dyad{\eta}).
    \end{equation}
    Assume that $\rho=\dyad{\psi}$, and $\ket{\phi}$ be a pure state orthogonal to $\ket{\psi}$. After some calculation, we have that
    \begin{equation}
        \Theta_{\dyad{\psi}}\circ (\Psi^{-1})^\ddag (\dyad{\phi})
    \end{equation}
    equals to
    \begin{equation}
        \bra{\psi}\ket{\eta}\bra{\eta}\ket{\phi} \dyad{\psi}{\phi}\qty(\frac{1}{1-r}\dyad{\eta}^\perp + \dyad{\eta}),
    \end{equation}
    which does not vanish when $\bra{\psi}\ket{\eta}\bra{\eta}\ket{\phi} \neq 0$. Choice of such a trio $\ket{\psi}, \ket{\phi}$ and $\ket{\eta}$ is possible in any $A$ with dimension higher than 1. This contradicts (\ref{eqn:axPinF}). In summary, there exists a state over time function that satisfies axioms \textbf{(E)}, \textbf{(QC)} and \textbf{(P)} as a reuslt, but not \textbf{(CC)}. Since axioms \textbf{(CC)} and \textbf{(J)} are equivalent under \textbf{(E)} and \textbf{(P)} by Proposition \ref{prop:ILPII}, such a function does not satisfy axiom \textbf{(J)} either.
\end{proof}
Finally, we show that axioms \textbf{(T)} and \textbf{(QC)} imply axiom \textbf{(J)}.
\begin{proof}
    In terms of state-rendering function, Axiom \textbf{(T)} is equivalent to $\Theta^\ddag_\rho:=\dag\circ\Theta^\dag_\rho\circ\dag=\Theta_\rho$, therefore we get $\Tr[\Theta_\rho(\cdot)]=\Tr[\mds{1}_A\Theta_\rho(\;\cdot\;)]=\Tr[\Theta^\ddag_\rho(\mds{1}_A)\;\cdot\;]=\Tr[\Theta_\rho(\mds{1}_A)\;\cdot\;]=\Tr[\rho\;\cdot\;]$ because $[\rho,\mds{1}_A]=0$ for all $\rho\in\mf{S}(A)$. Therefore we have from Eq. (\ref{eqn:marg})
    \begin{equation}
        \mcal{E}_{B|A}(\rho_A)=\Tr_A[\mcal{E}_{B| A}\star \rho_A ]
    \end{equation}
    and from Eq. (\ref{eqn:Qcond}),
    \begin{equation}
         \mcal{E}_{B|A}(\rho_A)=\Tr_A[(\rho_A\otimes \mds{1}_B)\qty(\mcal{E}_{B| A}\star \mds{1}_A )].
    \end{equation}
    This linear equation defined for $\rho\in\mf{S}(A)$ uniquely extends to $\mf{B}(A)$ (See Remark \ref{rmk:linear} in Appendix \ref{app:rmks}). Hence, we could let $A$ be a subsystem of a joint system $AE$. By letting $\rho_{AE}=F_{AE}$, the swap gate between $A$ and $E$, we have (after renaming $A$ to $A'$ and $E$ to $A$ in the end result)
    \begin{equation} \label{eqn:jamdec}
        \mscr{D}[\mcal{E}_{B|A}]=(\id_A\otimes \mcal{E}_{B|A'}) (F_{AA'})=\mcal{E}_{B| A}\star\mds{1}_A.
    \end{equation}
    This implies that $\mcal{E}_{B|A}\star\rho_A=(\id_A\otimes\mcal{E}_{B|A})(\id_{A'|A}\star\rho_A)$ where $\id_{A'}\star\rho_A=(\Theta_\rho\otimes\id_{A'})(F_{AA'})$.
\end{proof}

Because axiom \textbf{(P)} follows from \textbf{(E)} and \textbf{(QC)}, deriving axiom \textbf{(J)} implies axiom \textbf{(CC)} too, by Proposition \ref{prop:ILPII}.

\section{Specialness of \jami state} \label{app:nonU}

\begin{prop} \label{prop:nonU}
    A state over time function $\star$ satisfies axioms \textbf{(T)} (or \textbf{(H)}), \textbf{(E)}, \textbf{(P)} and \textbf{(\^{J})} if and only if it is induced through (\ref{eqn:Edecomp}) by an arbitrary time-expansion function $\id_{A'|A}\star\rho_A$ having the following expression,
\begin{equation} \label{eqn:general}
     \frac{1}{|A|} \qty(\Tr[\rho]F_{AA'} + \mds{1}_A \otimes \hat{\rho}_{A'} + \hat{\rho}_A \otimes \mds{1}_{A'}) + \Xi(\rho),
\end{equation}
where $\hat{\rho}:=\rho - (\Tr[\rho]/|A|) \mds{1}_A$ and $\Xi : \mf{B}(A) \to \mf{B}(AA')$ is an arbitrary linear function such that $\Xi(\mds{1}_A)=0$,  $F_{AA'}\Xi(\rho)F_{AA'}=\Xi(\rho)$ (or Hermitian-preserving) and $\Tr_A\circ\;\Xi=\Tr_{A'}\circ\; \Xi=0$.
\end{prop}
Proposition \ref{prop:nonU} shows that there are many state over time functions other than the FP function that satisfy all the axioms introduced im the main text except for \textbf{(CC)}. Moreover, if we replace $|A|^{-1}F_{AA'}$ with some operator $D_{AA'}$ with suitable properties, then the resultant state over time function still satisfies all the axioms \textbf{(T)} (or \textbf{(H)}), \textbf{(E)}, \textbf{(P)} and \textbf{(\^{J})}. This shows that, without the constraint on applicability to arbitrary subspaces, indeed the choice of the time-expansion function is independent of other axioms except for \textbf{(CC)} and can be made rather arbitrarily.

\begin{proof}
    We prove a more general result in which $|A|^{-1}F_{AA'}$ is replaced by an arbitrary operator $D_{AA'}$ on $AA'$ that is symmetric under swapping $A$ and $A'$ (or Hermitian) and satisfies the marginal conditions, i.e. $\Tr_A D_{AA'} =\pi_{A'}$ and $\Tr_{A'} D_{AA'} = \pi_A$, in both Eq. (\ref{eqn:noevent}) and (\ref{eqn:general}).

    Since the state over time is state-linear, by Corollary \ref{coro:TrEidE}, one can define the action of the star product on subsystem so that it satisfies axiom \textbf{(E)}.
    
    Since Eq. (\ref{eqn:general}) is symmetric under the exchange of $A$ and $B$ by assumptions on $D_{AA'}$ and $\Xi(\rho)_{AA'}$, axiom \textbf{(T)} holds. Axiom \textbf{(J)} can be derived by letting $\rho_A=\pi_A$, i.e. $\hat{\rho}_A=0$. Axiom \textbf{(P)} is the most nontrivial one, and we show it as follows.  We can re-express Eq. (\ref{eqn:general}) as follows.
    \begin{equation} \label{eqn:idstar}
        \Tr[\rho]\qty(D_{AA'}-2\pi_A\otimes \pi_{A'}) + \pi_A \otimes \rho_{A'} +  \rho_A\otimes \pi_{A'} + \Xi(\rho)_{AA'}.
    \end{equation}
    By definition, $\mcal{E}_{B|A}\star \rho_A$ can be obtained by applying $\id_A\otimes \mcal{E}_{B|A'}$ to (\ref{eqn:idstar}). In the same way, by noting that $\Tr_B[\mcal{E}_{B|A}\star\rho_A]=\rho_A$, $\id_{B'| B}\star(E_{B|A}\star\rho_A)$ can be calculated as
    \begin{gather}
        \rho_A \otimes (D_{BB'} - 2\pi_B \otimes \pi_{B'}) + \nonumber\\
        \pi_B\otimes(\mcal{E}_{B'|A}\star\rho_A)+(\mcal{E}_{B|A}\star\rho_A)\otimes\pi_{B'}+ \label{eqn:idEcomp}\\
        +(\id_A\otimes\Xi_{BB'})(\mcal{E}_{B|A}\star\rho_A) .\nonumber
    \end{gather}
    Again, by the definition of $\star$, $\mcal{F}_{C|B}\star(\mcal{E}_{B|A}\star\rho_A)$ can be obtained by acting $\id_{AB}\otimes F_{C|B'}$ on Eq. (\ref{eqn:idEcomp}). Note that taking the partial trace on $B$ of Eq. (\ref{eqn:idEcomp}) yields $\mcal{E}_{B'|A}\star\rho_A$. Therefore, applying $\id_A\otimes \mcal{F}_{C|B'}$ results in $(\mcal{F}\circ\mcal{E})_{C|A}\star\rho_A$ as $\star$ is assumed to obey Eq. (\ref{eqn:Edecomp}).

    Conversely, if a state over time function $\star$ satisfies axioms \textbf{(E), (P)} and \textbf{(J)}, then it is characterized by its time-expansion function $\rho_A\mapsto\id_{A'|A}\star\rho_A$ through Eq. (\ref{eqn:Edecomp}) by Proposition \ref{prop:prlin}. By Proposition \ref{prop:stlin}, it is also state-linear and hence can be uniquely extended to all operators on $A$. Recall that $\mf{B}(A)=\mds{C}\mds{1}_A\oplus \mf{T}(A)$ where $\mf{T}(A)$ is the space of all traceless operators in $\mf{B}(A)$. Moreover, by axiom \textbf{(J)}, it should be decomposed into
    \begin{equation}
        \Tr[\rho]D_{AA'}+\Phi(\hat{\rho}),
    \end{equation}
    with some linear map $\Phi$ from $\mf{T}(A)$ to $\mf{T}(AB)$.  Therefore, one could expand $\Phi(\sigma)$ as follows.
    \begin{equation}
        \frac{1}{|A|}\qty(\Tr_{A'}\Phi(\sigma)\otimes \mds{1}_{A'} +\mds{1}_A\otimes \Tr_A \Phi(\sigma)) + \Lambda(\sigma)
    \end{equation}
    with some $\Lambda(\sigma)\in\mf{T}(A)\otimes\mf{T}(A')$. Note that $\Tr[\Phi(\sigma)]=0$ for all $\sigma \in \mf{T}(A)$. Thus, by the marginality conditions, i.e. $\Tr_A[\id_{A'|A}\star\rho_A]=\rho_{A'}$ and $\Tr_{A'}[\id_{A'|A}\star\rho_A]=\rho_A$, we have
    \begin{equation}
        \Phi(\hat{\rho})=\frac{1}{|A|}\qty(\hat{\rho}_A\otimes \mds{1}_{A'}+\mds{1}_A\otimes \hat{\rho}_{A'})+\Lambda(\hat{\rho}),
    \end{equation}
    Now, we let $\Xi(\rho):=\Lambda(\hat{\rho})$. For axiom \textbf{(T)} to be satisfied, we first need $F_{AA'}D_{AA'}F_{AA'}=D_{AA'}$ for the case $\rho_A=\pi_A$. The term $|A|^{-1}\qty(\hat{\rho}_A\otimes \mds{1}_{A'}+\mds{1}_A\otimes \hat{\rho}_{A'})$ is evidently symmetric under the exchange of $A$ and $A'$. Thus, it follows that $\Xi(\hat{\rho})$ is also symmetric.

    Similarly, for axiom \textbf{(H)} to be satisfied, $D_{AA'}$ should be Hermitian and $\Xi(\rho)$ should be Hermitian for all $\rho\in\mf{S}(A)$ because $\hat{\rho}_A\otimes\mds{1}_{A'}+\mds{1}_A\otimes \hat{\rho}_{A'}$ is obviously Hermitian for all $\rho\in\mf{S}(A)$. Since $\mf{S}(A)$ spans $\mf{H}(A)$, it equivalent to $\Xi(\rho)$ being Hermitian-preserving.
\end{proof}
\section{Proof of Theorem \ref{thm:PF}} \label{App:B}
In light of Theorem \ref{thm:equiv}, we prove the uniqueness from axioms \textbf{(T)}, \textbf{(E)} and \textbf{(QC)}.
\begin{proof}
     Note that, by axiom \textbf{(T)}, we have $\Theta^\ddag_\rho:=\dag\circ\Theta^\dag_\rho\circ\dag=\Theta_\rho$. Interpret $\Theta_\rho(M)$ as a binary operation between $\rho$ and $M$, i.e. $\rho \odot M:=\Theta_\rho(M)$. Then, one could check that by the assumption of linearity in $\rho$, it follows that $\odot$ is bilinear. By the property of $\Theta_\rho$ in axiom \textbf{(QC)}, it also follows that $\rho \odot \mds{1}_A = \mds{1}_A \odot \rho = \rho$ because $[\rho,\mds{1}_A]=0$ for all $\rho\in\mf{S}(A)$ and that $\rho \odot M =0$ whenever $\rho$ and $M$ have orthogonal supports because that implies $[\rho,M]=0$ and $\rho M =0$. One also get $\Tr[\rho\odot M]=\Tr[\Theta_\rho(M)]=\Tr[(\Theta^\dag_\rho(\mds{1}_A))^\dag M]=\Tr[(\Theta^\dag_\rho(\mds{1}^\dag_A))^\dag M]=\Tr[\Theta_\rho(\mds{1}_A)M]=\Tr[\rho M]$. Since $\odot$ can be linearly extended to all $\rho\in\mf{B}(A)$, by Lemma 3 in Ref. \cite{heunen2013matrix}, it follows that
\begin{equation} \label{eqn:ASRgform}
    \Theta_\rho(M)= \rho M + g([\rho, M]),
\end{equation}
for some linear map $g$ on $\mf{B}(A)$ that maps traceless operators to traceless operators. Without loss of generality, we can set $g(\mds{1}_A)=\mds{1}_A$ because only its action on traceless operators matters in this proof. We remark that it is equivalent to $\Tr\circ \; g^\dag = \Tr$, i.e. $g^\dag$ is trace-preserving. When $\rho$ is a quantum state, $\Theta^\dddag_\rho$ has the following form.
\begin{equation}
    \Theta^\ddag_\rho(M)= M \rho -[\rho,  g^\ddag(M)],
\end{equation}
where $g^\ddag := \dag \circ g^\dag \circ \dag$. Note that we have $\Tr\circ g^\ddag =\Tr$, too. Therefore, the condition $\Theta^\ddag_\rho=\Theta_\rho$ implies that
\begin{equation} \label{eqn:AgMM2}
    g([\rho,M])+[\rho,M]=-[\rho, g^\ddag(M)],
\end{equation}
for all $\rho\in \mf{S}(A)$ and $M\in\mf{B}(A)$.
This equation linear in $\rho$ extends uniquely to all $\rho\in\mf{B}(A)$ (See Remark \ref{rmk:linear}). We have $[\dyad{\eta},g^\ddag(\dyad{\phi})]=0$ for any $\rho=\dyad{\eta}$ and $M=\dyad{\phi}$ with $\braket{\eta}=\braket{\phi}=1$ and $\bra{\eta}\ket{\phi}=0$ so that $[\dyad{\eta},\dyad{\phi}]=0$. By Schur's lemma \cite{schur1905neue}, it follows that
\begin{equation} \label{eqn:Agphi}
    g^\ddag(\dyad{\phi})=c_\phi \dyad{\phi}+ d_\phi (\mds{1}_A-\dyad{\phi}), 
\end{equation}
for some complex numbers $c_\phi$ and $d_\phi$. Now, we observe that, because of the trace-preserving property of $g^\ddag$, by taking the trace of Eq. (\ref{eqn:Agphi}), we have
\begin{equation} \label{eqn:Ac+d}
    1=c_\phi+d_\phi(|A|-1),
\end{equation}
for any unit vector $\ket{\phi}\in A$.

On the other hand, note that $g([\rho, M])+[g,M]=-g([M, \rho])-[M,g]$. Hence, by applying Eq. (\ref{eqn:AgMM2}) with $\rho$ and $M$ swapped, we get
\begin{equation}\label{eqn:AgMMg}
    [\rho, g^\ddag(M)] = [g^\ddag (\rho), M],
\end{equation}
for all $\rho, M \in \mf{B}(A)$. If $[\dyad{\phi},\dyad{\psi}]\neq 0$, by setting $\rho=\dyad{\phi}$ and $M=\dyad{\psi}$, we have from Eq. (\ref{eqn:Agphi}) and Eq. (\ref{eqn:AgMMg}) that
\begin{equation}
    (c_\phi - d_\phi) [\dyad{\phi},\dyad{\psi}]=(c_\psi - d_\psi) [\dyad{\phi},\dyad{\psi}].
\end{equation}
Since we assumed that $[\dyad{\phi},\dyad{\psi}]\neq 0$, we get
\begin{equation} \label{eqn:Ac-d}
    c_\phi - d_\phi = c_\psi - d_\psi.
\end{equation}
In the case $[\dyad{\phi},\dyad{\psi}]=0$ but $\dyad{\phi}\neq\dyad{\phi}$, by using $\ket{\nu}:=2^{-1/2}(\ket{\phi}+\ket{\psi})$, which is not orthogonal to either $\ket{\phi}$ or $\ket{\psi}$, we can show that
\begin{equation}
    c_\phi-d_\phi=c_\nu-d_\nu=c_\psi-d_\psi.
\end{equation}
In other words, $c_\phi-d_\phi$ is independent of $\ket{\phi}\in A$. Combining it with Eq. (\ref{eqn:Ac+d}), we obtain that $c:=c_\phi$ and $d:=d_\phi$ are also independent of unit vector $\ket{\phi}\in A$. From linearity of $g^\ddag$, it follows that for every $M\in \mf{B}(A)$,
\begin{equation}
    g^\ddag(M)=d\Tr[M]\mds{1}_A+(1-d|A|)M.
\end{equation}
Equivalently, for every $N\in \mf{B}(A)$,
\begin{equation}
    g(N)=d\Tr[N]\mds{1}_A+(1-d|A|)N.
\end{equation}
Using this expression in Eq. (\ref{eqn:ASRgform}) results in
\begin{equation} \label{eqn:Amu}
    \Theta_\rho(M)= \mu \rho M + (1-\mu) M \rho,
\end{equation}
by letting $\mu:=2-d|A|$. Finally, by observing that
\begin{equation} \label{eqn:Amu2}
    \Theta^\ddag_\rho(M)= (1-\mu) \rho M + \mu M \rho,
\end{equation}
we get $\mu=1/2$ from $\Theta_\rho=\Theta^\ddag_\rho$.
\end{proof}

\section{Proof of Proposition \ref{prop:PFuni}} \label{app:PFuni}

\begin{proof}
    By assumption, $\Theta_\rho$ is self-adjoint, i.e. $\Theta^\dag_\rho=\Theta_\rho$. By following a similar line of argument with the proof of Theorem \ref{thm:PF} in Appendix \ref{App:B}, we can see that the binary operation $\rho\odot M:=\Theta_\rho(M)$ satisfies all the properties therein, hence, again by Lemma 3 of Ref. \cite{heunen2013matrix}, we have
\begin{equation} \label{eqn:SRgform}
    \Theta_\rho(M)= \rho M + g(\rho M - M \rho),
\end{equation}
for some linear map $g$ on $\mf{B}(A)$ that maps traceless operators to traceless operators. Without loss of generality, we can set $g(\mds{1}_A)=\mds{1}_A$ because only its action on traceless operators matters in this proof. We remark that it is equivalent to $\Tr\circ \; g^\dag = \Tr$, i.e. $g^\dag$ is trace-preserving. When $\rho$ is a quantum state, the adjoint of $\Theta_\rho$ has the following form.
\begin{equation}
    \Theta^\dag_\rho(M)=\rho M + \rho g^\dag(M) - g^\dag (M) \rho.
\end{equation}
Therefore, the condition $\Theta^\dag_\rho=\Theta_\rho$ implies that
\begin{equation} \label{eqn:gMM}
    g(\rho M - M \rho)=\rho g^\dag(M) - g^\dag (M) \rho,
\end{equation}
for all $\rho\in \mf{S}(A)$ and $M\in\mf{B}(A)$. In other words,
\begin{equation} \label{eqn:gMM2}
    g([\rho,M])=[\rho, g^\dag(M)],
\end{equation}
This equation linear in $\rho$ extends uniquely to all $\rho\in\mf{B}(A)$ (See Remark \ref{rmk:linear}). We have $[\dyad{\eta},g^\dag(\dyad{\phi})]=0$ for any $\rho=\dyad{\eta}$ and $M=\dyad{\phi}$ with $\braket{\eta}=\braket{\phi}=1$ and $\bra{\eta}\ket{\phi}=0$. Therefore, we can apply exactly the argument used in the proof of Theorem \ref{thm:PF} to $g^\dag$ instead of $g^\ddag$ and get the following result: For every $M\in \mf{B}(A)$,
\begin{equation}
    g^\dag(M)=d\Tr[M]\mds{1}_A+(1-d|A|)M.
\end{equation}
Equivalently, for every $N\in \mf{B}(A)$,
\begin{equation}
    g(N)=d^*\Tr[N]\mds{1}_A+(1-d^*|A|)N.
\end{equation}
Using this expression in Eq. (\ref{eqn:SRgform}) results in
\begin{equation} \label{eqn:mu}
    \Theta_\rho(M)= \mu \rho M + (1-\mu) M \rho,
\end{equation}
for some coefficient $\mu  \in \mds{R}$. Note that $\mu$ being real follows from $\Theta_\rho=\Theta^\dag_\rho$.

Requiring $\Tr[M^\dag \Theta_\rho(M)] \geq 0$ for all $M\in \mf{B}(A)$ whenever $\rho \in \mf{S}(A)$, the coefficient $\mu$ in (\ref{eqn:mu}) must satisfy $0\leq \mu \leq 1$. This can be seen in the cases in which $\rho=\dyad{\phi}$ and $M=\dyad{\phi}{\psi}$ or $M=\dyad{\psi}{\phi}$ with $\braket{\phi}{\psi}=0$.  This makes $\Theta_\rho$ a convex sum of the left bloom $\Theta^L_\rho(M):= M \rho$ and the right bloom $\Theta^R_\rho(M):= \rho M$.
\end{proof}

\section{Insufficiency of axiom \textbf{(H)}} \label{app:insuff}

\begin{prop}
    For any real number $c$, the state over time function $\star$ induced by the state-rendering function $\Theta_\rho$ given as
    \begin{equation} \label{eqn:ahsrf}
        \Theta_\rho(M)=\frac{1}{2}\qty{\rho,M} + ic[\rho,M],
    \end{equation}
    through $\mcal{E}_{B|A}\star\rho_A:=(\Theta_\rho\otimes\id_B)(\mscr{D}[\mcal{E}_{B|A}])$,
    satisfies axioms \textbf{(H)} and \textbf{(E)}, \textbf{(P)}, \textbf{(CC)}, \textbf{(J)} and \textbf{(QC)}.
\end{prop}
\begin{proof}
    Since $\Theta_\rho$ is Hermitian-preserving and $\mscr{D}[\mcal{E}_{B|A}]$ is Hermitian for any quantum channel $\mcal{E}_{B|A}$, axiom \textbf{(H)} is satisfied. Also, as $\Theta_\rho$ is linear in $\rho$, the induced state over time function is state-linear, so axiom \textbf{(E)} is satisfied by Proposition \ref{prop:stlin}. Axiom \textbf{(P)} follows from that $\Theta_\rho$ is linear in $\rho$ and that the induced state over time function is defined through Eq. (\ref{eqn:Edecomp}) by Proposition \ref{prop:prlin}.

     Next, whenever $A=\bigoplus_i A_i$ and $\rho\in\mf{S}(A)$ is decomposed into $\rho=\sum_i \lambda_i \rho_i$ with some probability distribution $\qty{\lambda_i}$ and states $\rho_i\in\mf{S}(A_i)$, for any $M_i\in\mf{B}(A_i)$, one can observe that
     \begin{equation}
         \Theta_\rho\qty(\sum_i M_i)= \sum_i \lambda_i \Theta_{\rho_i}(M_i),
     \end{equation}
     because
     \begin{equation}
         \qty{\sum_i \lambda_i \rho_i, \sum_i M_i}=\sum_i \lambda_i \qty{\rho_i,M_i}
     \end{equation}
     and similarly
     \begin{equation}
         \qty[\sum_i \lambda_i \rho_i, \sum_i M_i]=\sum_i \lambda_i \qty[\rho_i,M_i].
     \end{equation}
     Therefore, it follows that axiom \textbf{(CC)} is also satisfied. Axiom \textbf{(J)} follows due to Proposition \ref{prop:ILPII}. One can easily check that axiom \textbf{(QC)} is satisfied by observing that $\Theta_\rho(M)$ is linear in $\rho$ and reduces to $\rho M$ when $[\rho,M]=0$. See Ref. \cite{parzygnat2023virtual} for a related discussion on this one-parameter family of quantum states over time.

\end{proof}

\bibliography{main}
\end{document}